\documentclass[12pt]{article}

\usepackage{verbatim}
\usepackage{amsmath}
\usepackage{amssymb}
\usepackage{graphicx}
\usepackage{cite}
\usepackage{subfig}
\usepackage{setspace}
\usepackage{url}
\usepackage[percent]{overpic}
\usepackage{slashed}
\usepackage{authblk}

\usepackage{xspace}
\newcommand{\Fermi}[0]{\textit{Fermi}\xspace}

\usepackage{fullpage}
\usepackage{hyperref}
\def\tev{\,{\rm TeV}}
\def\gev{\,{\rm GeV}}
\def\ie{{\it i.e.}}
\def\eg{{\it e.g.}}

\def\Fermi{\,{\it Fermi}}

\title{Dark Matter Complementarity in the pMSSM and the ILC{\footnote{Talk presented at the International Workshop on Future 
Linear Colliders(LCWS13), Tokyo, Japan, 11-15 November 2013}}}
\date{}
\author[1]{Thomas G. Rizzo}
\affil[1]{SLAC National Accelerator Laboratory, Menlo Park, CA, 94025, USA\footnote{rizzo@slac.stanford.edu}}

\begin{document}

\rightline{\vbox{\halign{&#\hfil\cr
&SLAC-PUB-15887\cr
}}}


{\let\newpage\relax\maketitle}

\begin{abstract}
The search for and potential identification of dark matter (DM) will require a simultaneous, multi-pronged approach with important roles 
to be played by the LHC, both direct and indirect DM detection experiments and, eventually, the ILC. The 19-parameter p(henomenological)MSSM 
can provide a broad framework for complementarity studies of neutralino DM. In this talk, I summarized the sensitivity of these searches 
at the 7, 8 and 14 TeV LHC, combined with those from \Fermi, CTA, IceCube/DeepCore, COUPP, LZ and XENON. The strengths and weaknesses of 
each of these searches is briefly discussed as is their interdependent roles in probing the pMSSM model parameter space. I also comment 
on the future role of the ILC in exploring the detailed nature of neutralino DM. 
\end{abstract}

\section{Introduction and Overview of Complementarity}

Although the reality of DM is no longer in question, its true nature remains as mysterious as ever. Because of a 
numerical `miracle', the simplest and most often studied possibility is that DM is a Weakly Interacting Massive Particle (WIMP) whose 
thermal relic abundance is set by its annihilation cross section in the early universe. Beyond simplicity, one of the reasons WIMPs are 
so attractive is that they can be sought using multiple experimental techniques: through direct production at accelerators, direct detection 
(DD) by scattering off nuclei in experiments underground and indirect detection (ID) of DM annihilation products coming from the sky. Even 
if convincing DM signals are discovered in one of these channels in the future, however, it is likely that complementary information from 
all three of these sources will be necessary to elucidate the detailed properties of DM. So far, no such conclusive signals for DM have 
yet been observed and the parameter space for WIMPs continues to slowly shrink though much yet remains to be explored. 

In order to make use of possible complementary information on the properties of DM one must have a framework that completely describes 
how the DM interacts with the various particles of the Standard Model (SM). Here there are two different approaches: in a bottom up, 
`model-independent' framework, one posits a simplified model with only the minimal particle content, \ie, the DM particle itself as well as 
a mediator connecting the DM with the SM.  Furthermore, provided the mediator is itself sufficiently massive to be integrated out, the 
resulting interactions are described by a set of higher dimension operators, \ie, an EFT. In the situation where the EFT is applicable 
this approach has been shown to be quite powerful, especially for DD and ID searches. However, in the case of colliders such as the LHC this 
EFT approach frequently breaks down and one needs to revert back to the original simplified model for consistency. Of course if DM is not the 
only light(ish) state even this simplified model approach will itself fail. A second possibility is to adopt a UV-complete scenario where the 
DM particle is accompanied by a potentially wide range of other new states of various masses all of which can interact with the SM through 
electroweak and/or strong interactions. Unlike in the EFT or simplified model approaches, in this case we are likely forced to buy into a 
rather complex scenario with a multi-dimensional parameter space that needs to be examined.  

Of the possible UV-complete scenarios predicting WIMPs, R-parity conserving supersymmetry (SUSY) appears to be the most attractive with the 
MSSM being its simplest realization. Here the lightest stable SUSY particle (LSP) that is both electrically neutral and a color-singlet, often 
the lightest neutralino ($\chi_1^0$) as we will assume in the discussion below, can be identified as a possible WIMP DM candidate. While DM 
searches are obviously focused on the properties of the LSP itself, within this UV framework the properties of many if not all of the other 
superparticles as well as those of the members of the extended SUSY Higgs sector can also play important roles in the various searches. Thus 
it is impossible to completely separate DM searches from searches for and the examination of the rest of the SUSY spectrum. Furthermore, even 
in the simplest SUSY scenario, the MSSM, the number of free parameters ($\sim$ 100) is simply too large to perform a study in all generality 
so we need to restrict our attention in some way without (hopefully) losing contact with any of the relevant physics. 

To at least partially circumvent this problem, in the analysis discussed here we restrict ourselves to the 19-parameter p(henomenological)MSSM 
as is fully discussed detail elsewhere{\cite {us}}. For completeness, the 19 pMSSM parameters and their ranges employed in the results below 
are presented in Table~\ref{ScanRanges}. To study the pMSSM, we essentially throw darts into this large parameter space, generating many 
millions of random model points, with each point corresponding to a specific set of values for these 19 parameters. These individual `models' 
are then subjected to a large set of collider, flavor, precision measurement, dark matter and theoretical constraints with roughly 225k models 
surviving these for further study here. One interesting result of employing these scan ranges is that the LSP is frequently (but not always) 
close to being an electroweak eigenstate, \ie, a relatively pure bino, wino or Higgsino, which has important consequences for DM searches. 

Below, we will briefly overview the sensitivities of both present and future experiments which employ different approaches in searching for 
SUSY dark matter: collider production, as well as direct and indirect detection, to see how well this pMSSM parameter space is `covered'. The 
analyses will {\it not} assume that the thermal relic density as calculated for the LSP necessarily saturates the WMAP/Planck value, 
$\Omega h^2\simeq 0.12$, to allow for the possibility of multi-component DM. For example, the axion introduced to solve the strong CP problem 
might make up a substantial amount of the DM. In the pMSSM the nature of the neutralino LSP can vary widely as one moves about in this large 
parameter space. Even if the LSP is is (in)directly observed at the LHC it is likely to be difficult to fully determine its electroweak 
properties even with the supplementary information provided by the DD and/or ID DM experiments. These properties are of course critical to 
identifying the LSP as a true component of DM. It is likely that this information can only be provided by the ILC, which can fully study the 
nature of the LSP provided it is kinematically accessible, in order to complete a consistent picture of DM.

\begin{table}
\centering
\begin{tabular}{|c|c|} \hline\hline
$m_{\tilde L(e)_{1,2,3}}$ & $100 \gev - 4 \tev$ \\ 
$m_{\tilde Q(q)_{1,2}}$ & $400 \gev - 4 \tev$ \\ 
$m_{\tilde Q(q)_{3}}$ &  $200 \gev - 4 \tev$ \\
$|M_1|$ & $50 \gev - 4 \tev$ \\
$|M_2|$ & $100 \gev - 4 \tev$ \\
$|\mu|$ & $100 \gev - 4 \tev$ \\ 
$M_3$ & $400 \gev - 4 \tev$ \\ 
$|A_{t,b,\tau}|$ & $0 \gev - 4 \tev$ \\ 
$M_A$ & $100 \gev - 4 \tev$ \\ 
$\tan \beta$ & 1 - 60 \\
\hline\hline
\end{tabular}
\caption{Scan ranges for the 19 parameters of the pMSSM with a neutralino LSP employed in the present analysis. All parameters are scanned with 
flat priors~\cite{us}.}
\label{ScanRanges}
\end{table}

\begin{figure}[htbp]
\centerline{\includegraphics[width=7.0in]{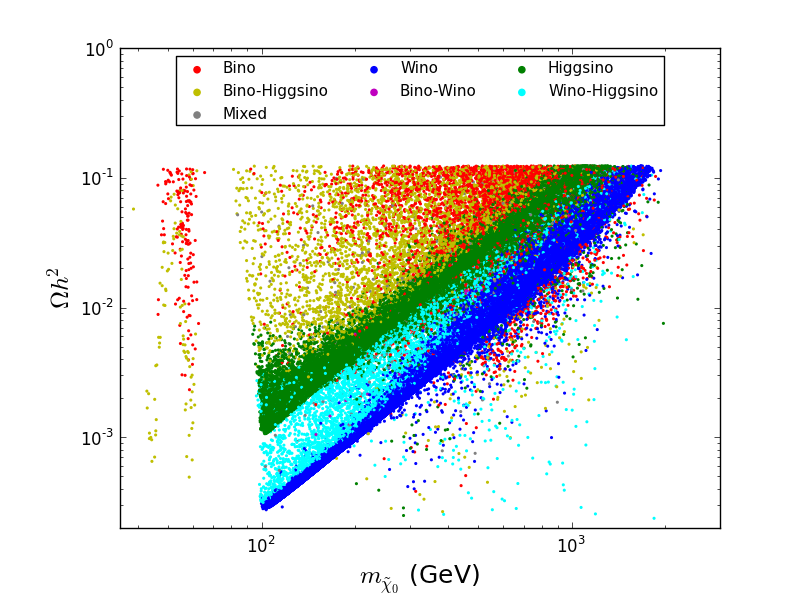}}
\vspace*{-0.10cm}
\caption{Thermal relic density as a function of the LSP mass in our pMSSM model set, as generated, color-coded by the electroweak properties of the  
LSP as discussed in the text.}
\label{fig00}
\end{figure}

Figure~\ref{fig00} provides an overview of the properties of the many LSPs in our model set. Here we see the thermal relic densities of the LSPs 
as a function of their mass with the color-coding reflecting their electroweak eigenstate content. There are many things to note here that will be 
important for later consideration. Essentially every possible mechanism to obtain (or to lie below) the WMAP/Planck relic density is present here: 
($i$) The set of models with low LSP masses (forming `columns' on the left-hand side of the Figure) correspond to bino-Higgsino admixtures which 
achieve a sufficiently low relic density by resonant annihilation through the $Z,h$-funnels; these sometimes appear as pure binos if the Higgsino 
fraction is very small{\footnote {Here, `pure' means having an eigenstate fraction $\geq 90\%$. Points shown as bino-wino, bino-Higgsino, or 
wino-Higgsino mixtures have less than $2\%$ Higgsino, wino, or bino fraction, respectively. Mixed points have no more than $90\%$ and no less than 
$2\%$ of each component.}}. ($ii$) The bino-Higgsino LSPs saturating the relic density in the upper left region of the figure are 
of the so-called `well-tempered' variety. ($iii$) The pure bino models in the upper middle region of the Figure are bino co-annihilators (mostly 
but not exclusively with with sleptons) or those which annihilate resonantly through the $A$-funnel. ($iv$) The green (blue) bands are pure 
Higgsino (wino) models that saturate the relic density bound (using perturbative calculations which do not include Sommerfeld enhancement) near 
$\sim 1(1.7)$ TeV and have very low relic densities for lighter LSP masses. Wino-Higgsino hybrids are seen to lie between these two cases as 
expected. ($v$) A smattering of models with additional (or possibly multiple) aids to their annihilation are loosely distributed in the lower 
right-hand corner of the Figure. As we will see, many of the searches for DM are particularly sensitive to one or more of these LSP categories. 
We now turn to a brief discussion of the individual searches.

\section{DM Searches}

\subsection{LHC}

The first set of constraints on the neutralino LSP DM scenario arise from the direct searches for SUSY sparticle production at the LHC at both 
7 and 8 TeV. In general, our approach here is to replicate the suite of ATLAS SUSY analyses employing fast Monte Carlo but also to supplement 
these with several searches performed by CMS; the analyses we consider are briefly summarized in Table~2 of Ref.~\cite {us} and are validated 
using both ATLAS and CMS benchmark model points. These sets of searches play distinct but important roles in restricting the pMSSM parameter 
space. The details of our procedure can be found in Ref.{\cite {us}} and here we simply display some of the results. Interestingly, we find that, 
once combined, the total fraction of our models surviving (or killed by) the set of all LHC searches is to an excellent approximation 
{\it independently} of whether or not the Higgs mass constraint (\ie, requiring $m_h=126\pm 3$ GeV) has been applied.

\begin{figure}[htbp]
\centerline{\includegraphics[width=3.5in]{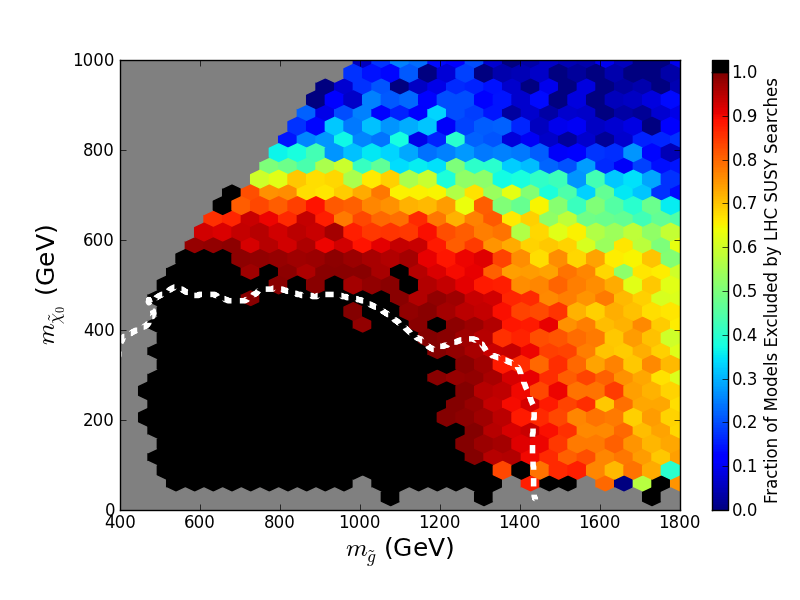}
\hspace{-0.50cm}
\includegraphics[width=3.5in]{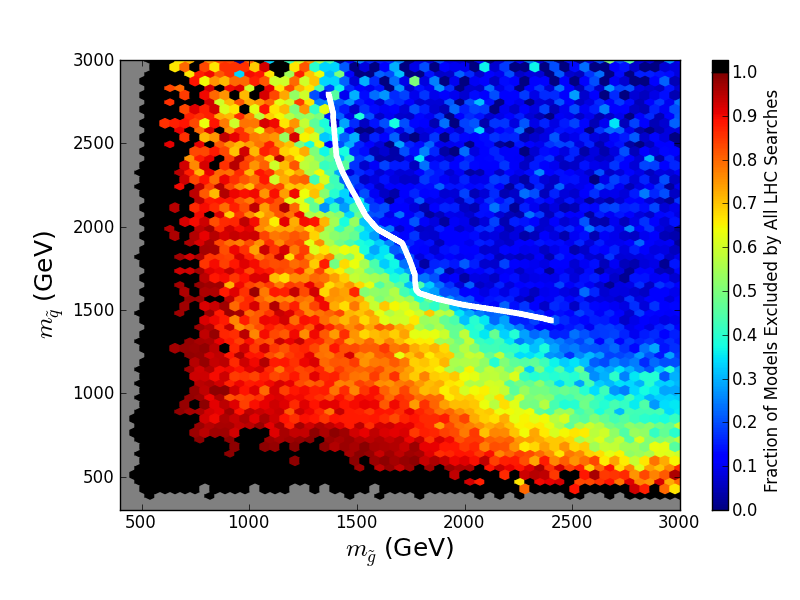}}
\vspace*{0.50cm}
\centerline{\includegraphics[width=3.5in]{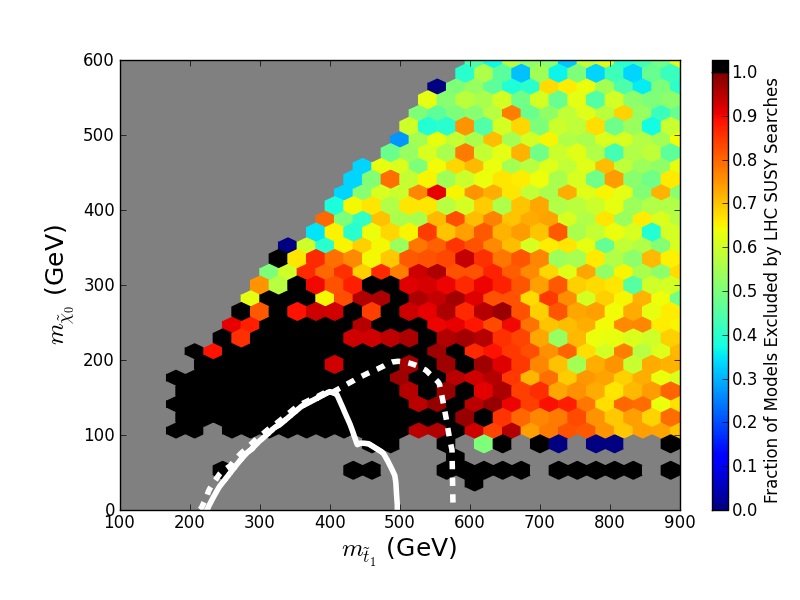}
\hspace{-0.50cm}
\includegraphics[width=3.5in]{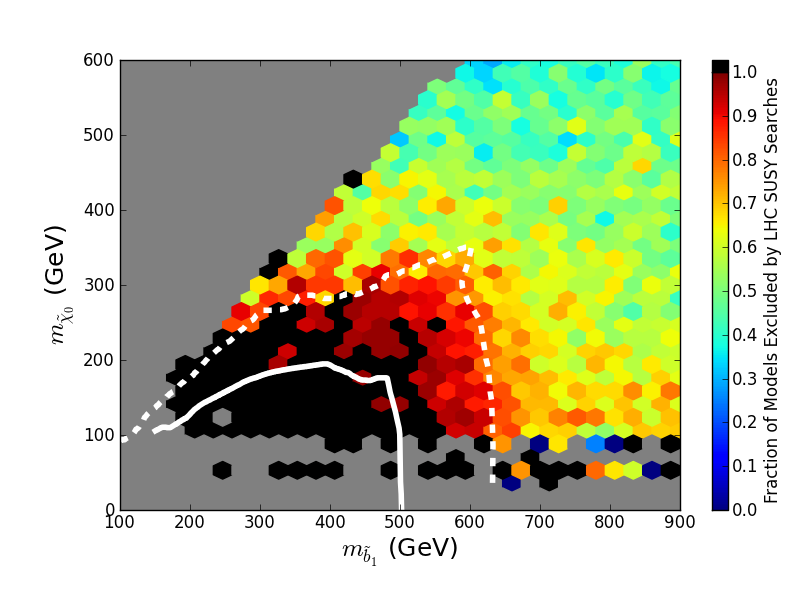}}
\vspace*{-0.10cm}
\caption{Projections of the pMSSM model coverage efficiencies from the 7 and 8 TeV LHC searches shown in the gluino-LSP mass plane (top left), 
the gluino-lightest squark mass plane (top right), the lightest stop-LSP mass plane (lower left) and in the lightest sbottom-LSP mass plane. 
The solid/dashed lines represent the corresponding $95\%$ CL limit results obtained by ATLAS in the simplified model context.}
\label{fig1}
\end{figure}

Figure~\ref{fig1} displays the fraction of models having a given sparticle and/or LSP mass that are excluded by the combined LHC searches. 
Due to their larger cross sections these searches are primarily focused on the production of colored sparticles that subsequently cascade 
decay down to the LSP. In the upper left-hand panel the coverage of the gluino-LSP mass plane by these searches is displayed; the white line 
represents the $95\%$ CL search limit on a simplified model with a gluino NLSP, neutralino LSP, and all other sparticles decoupled, obtained 
by ATLAS from their $\sim 20$ fb$^{-1}$ 2-6 jets plus MET analysis, which we see is roughly the same as the all-black region excluded in the 
pMSSM. The upper right-hand panel shows the corresponding coverage in the gluino-lightest squark mass plane with the corresponding simplified 
model line from ATLAS, here assuming that the LSP is massless and that the 8 squarks of the first two generations are all degenerate. As a 
result, since this is not true in the pMSSM, it is no surprise that our excluded region is not well described by the simplified model; while 
most models with rather light squarks and/or gluinos are observed to be excluded by the combined LHC searches, it is clear that models with 
squarks and/or gluinos below $\sim 700-750$ GeV still remain viable. The lower two panels show the lightest stop/sbottom vs. LSP mass plane 
with the corresponding ATLAS simplified model limits. Here we see several things, the most important being that the amount of coverage in these 
two mass planes in the pMSSM by the LHC searches is significantly larger than what we would have expected from either simplified model analysis. 
Clearly, the third-generation searches cover important gaps in the standard jets + MET coverage and vice-versa. In general, the third-generation 
searches are most sensitive to models in which stop or sbottom decays produce hard $b$-jets, and have a more difficult time observing models 
with stops or sbottoms that produce tops or decays to multiple intermediate gauginos (resulting in softer $b$-jets and leptons). The direct 
sbottom search is found to be responsible for the much of the heavy flavor search exclusion power for models with relatively light stops.

\subsection{Direct Detection}

The direct detection of DM results from the spin-independent (SI) and/or spin-dependent (SD) scattering of the LSP by a target nucleus. 
While $Z(h)$ $t-$channel exchange only contributes to the SD (SI) process at tree level, $s-$channel squark exchange can contribute to both 
processes. However, as the lower bounds on the first and second generation squark masses from the LHC become stronger the importance of these 
squark exchange contributions clearly will become subdominant.{\footnote {It is important to note, however, that since the many different 
light squark masses can vary independently in the pMSSM, and since the LHC constraints on, e.g., the $u_L,d_L$, $u_R$ and $d_R$ squarks are 
quite different, both SD and SI interactions may have substantial isospin-dependent contributions so that $\sigma_p$ and $\sigma_n$ can be 
significantly different.}} The $Z$-exchange graph is sensitive to the Higgsino content of the LSP whereas the Higgs exchange graph probes the 
product of the LSP's gaugino and Higgsino content. Similarly, (valence) squark exchange is particularly sensitive to the LSP's wino and bino 
content. 

\begin{figure}[htbp]
\centerline{\includegraphics[width=3.5in]{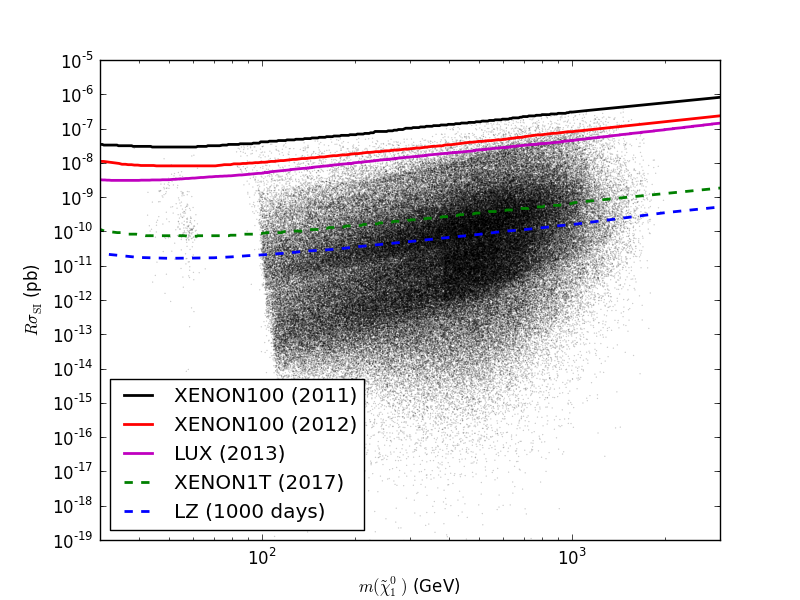}
\hspace{-0.50cm}
\includegraphics[width=3.5in]{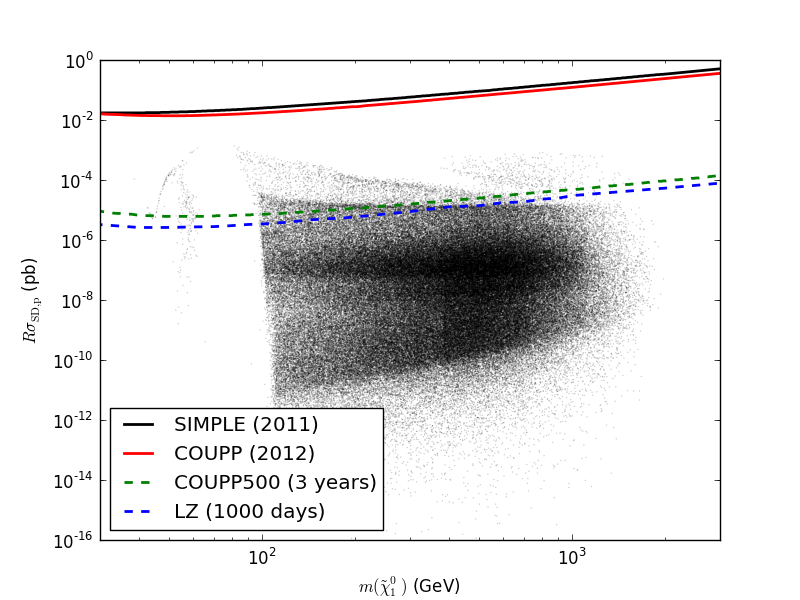}}
\vspace*{-0.10cm}
\caption{Scaled spin-independent (left) and spin-dependent (right) direct detection cross sections for our neutralino LSPs in comparison to current 
and future experimental sensitivities. The scaling factor accounts for the possibility that the calculated thermal relic density of the LSP is below 
that measured by WMAP.}
\label{fig3}
\end{figure}

Figure~\ref{fig3} displays the predicted experimentally observable SI and SD cross sections for this pMSSM model set together with several 
present and anticipated future experimental constraints. Note that the actual cross sections are appropriately scaled by the factor 
$R=\Omega h^2/\Omega h^2|_{WMAP/Planck}$ which accounts for the fact that most of our pMSSM models lead to thermal relic densities somewhat 
below that which saturates the WMAP/Planck value as discussed above. There are several things to note in this Figure: ($i$) Future SI searches 
will cut rather deeply into the model set; a lack of signal at XENON1T (LZ) would exclude 23\% (39\%) of these models. Increased data taking 
by LZ may enhanced this coverage to $\sim 48\%$ before potemtially running into the neutrino background. {\it However}, this implies that about  
half of our models are not accessible to SI experiments due to their rather small scaled cross sections. Models with these values of $R\sigma$ 
tend to produce these small values due to both the suppression arising from their low thermal relic density as well as the tendency of LSPs to 
be nearly pure weak eigenstates, as discussed above, which effectively turns off the tree-level $t-$channel exchanges. Note that if we only 
consider models which predict a relic density within 10\% of the measured value, the situation improves significantly, with 60\% (81\%)) of 
the models within the XENON1T (spin-independent LZ) reach. ($ii$) SD measurements are still rather far from the pMSSM model predictions and 
we do not expect future SD measurements to have a significant impact on the parameter space except in the low mass range. SD experiments such 
as COUPP500 (LZ) will only be able to exclude only $\sim 2(4)\%$ of the models in this set if no signal is seen.

\subsection{Indirect Detection: \Fermi ~LAT, CTA and IceCube}

Indirect detection can play a critical role in searches for DM and, in the case of null results, can also lead to very strong constraints 
on the pMSSM parameter space. As is well-known {\cite {us}}, searches for excess photons by both \Fermi (from, \eg, dwarf galaxies) ~and 
CTA (from, \eg, the galactic core) can contribute coverage in different regions of the pMSSM parameter space in the future.  CTA, in 
particular, is found to be extremely powerful in the search for heavy LSPs which are mostly Higgsino- or wino-like and that predict thermal 
relic densities within an order of magnitude or so of the WMAP/Planck value; these constitute $\sim 20\%$ of the present pMSSM model set. 
This role is of particular importance since these heavy LSPs are currently outside the range of the 7 and 8 TeV LHC SUSY searches and might 
be difficult to directly access even at 14 TeV. \Fermi, ~on the other hand, is found to be mostly sensitive to the set of well-tempered 
neutralinos that are relatively light. In the results presented below, the relevant analyses were performed by the \Fermi and CTA collaborations 
themselves{\cite {us}} and required both the calculation of the photon fluxes for each of the pMSSM model points under consideration as well 
as the corresponding modeling of the detector response in order to compare to their expected DM search sensitivities. A complete discussion 
of the procedures followed in this analysis and the corresponding details of these results can be found in Ref.~\cite {us}. 

In addition to these searches, IceCube/DeepCore can also make an important contribution to the pMSSM parameter space coverage. Neutralino 
dark matter can be captured, accumulate in the solar core and then annihilate into SM particles which subsequently decay to $\sim$ GeV 
neutrinos which may then be observable on Earth by IceCube. The observable flux for this is a result of the balance between the LSP capture 
rate in the sun and the corresponding DM annihilation rate. If the product of capture and annihilation cross sections is large enough this 
process leads to an equilibrium population of captured neutralinos whose annihilations are proportional to their elastic scattering cross 
sections.  However, in our pMSSM model, set $\sim 50\%$ of the models do not lead to such an equilibrium state which then results in a 
reduction of the effectiveness of the corresponding IceCube searches for neutrinos from LSP annihilations. As was the case for both the CTA 
and \Fermi analyses, the resulting flux needed to be calculated for each individual model point as LSP annihilation usually results in a 
weighted superposition of many SM final states.  IceCube, similar to \Fermi, is found to be most sensitive to Higgsino-bino admixtures 
which lie in the mass range below $\sim 700$ GeV. The details of this analysis can be found in Ref.~\cite{us} and here we make use only of 
the corresponding results.

\section{Complementarity: Combining the Searches}

\begin{figure}[htbp]
\centerline{\includegraphics[width=3.5in]{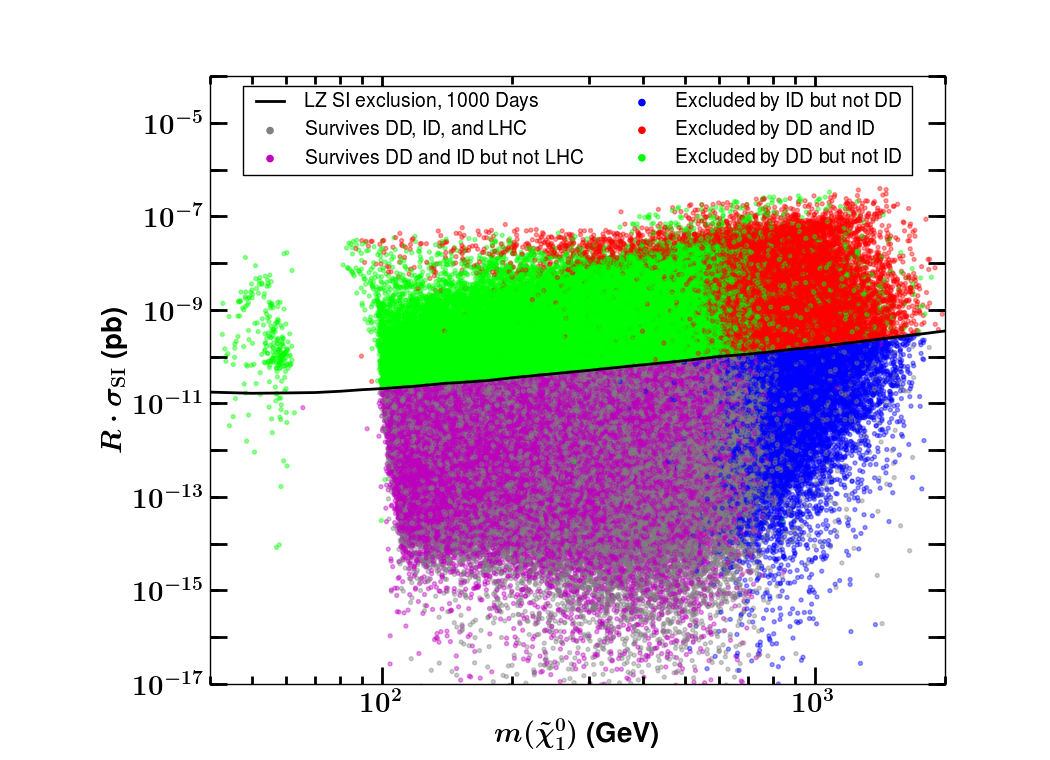}
\hspace{-0.50cm}
\includegraphics[width=3.5in]{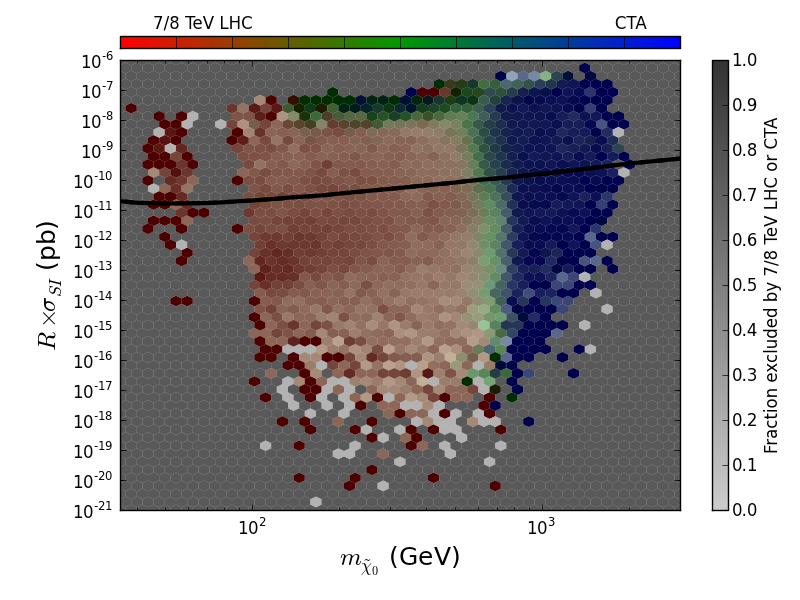}}
\vspace*{-0.10cm}
\caption{Comparisons of the models surviving or being excluded by the various searches in the LSP mass-scaled SI cross section plane as discussed 
in the text. The LZ SI line is shown as a guide to the eye.}
\label{figxx}
\end{figure}

Given all the pieces, we can now put them together to see what they tell us about the nature of the neutralino LSP as DM and, more 
generally, the pMSSM itself. Fig.~\ref{figxx} shows the survival and exclusion rates resulting from the various searches and their 
combinations in the LSP mass-scaled SI cross section plane. The left panel compares these for the combined direct detection 
(DD = SI + SD LZ) and indirect detection (ID = \Fermi ~+ CTA) DM searches with current results from the LHC. Here we see that 7.2\% (27.5\%) 
of the models are excluded by ID but not DD (excluded by DD but not ID) while 11.4\% are excluded by both searches. We also see that 
53.9\% of the models survive both sets of DM searches; 45.5\% of this subset of models, in turn, are presently excluded by the LHC searches. 
Note that the DD- and ID-excluded regions are all relatively well separated in terms of mass and cross section although there is some 
overlap between the sets of models excluded by the different experiments. In particular we see that the ID searches (here almost entirely 
CTA) are covering the heavy LSP region even in cases where the SI cross section is very low and likely beyond the reach of any potential 
DD experiment due to neutrino backgrounds. Combining all of the searches only 24.7\% of the present model set would survive in the absence 
of a signal. In the right panel the relative contributions arising from the LHC and CTA searches to the model survival/exclusion are shown. 
Here the intensity of a given bin indicates the fraction of models excluded there by the combination of both searches while the hue indicates 
whether the excluded models are seen mostly by CTA (blue) or by the LHC (red). It is again quite clear that CTA completely dominates for 
large LSP masses (which mostly correspond to wino and Higgsino LSPs) and also competes with the LHC throughout the band along the top of the 
distribution, which mostly contains models with thermal relic densities that approximately saturate the WMAP/Planck result.

\begin{figure}[htbp]
\centerline{\includegraphics[width=3.5in]{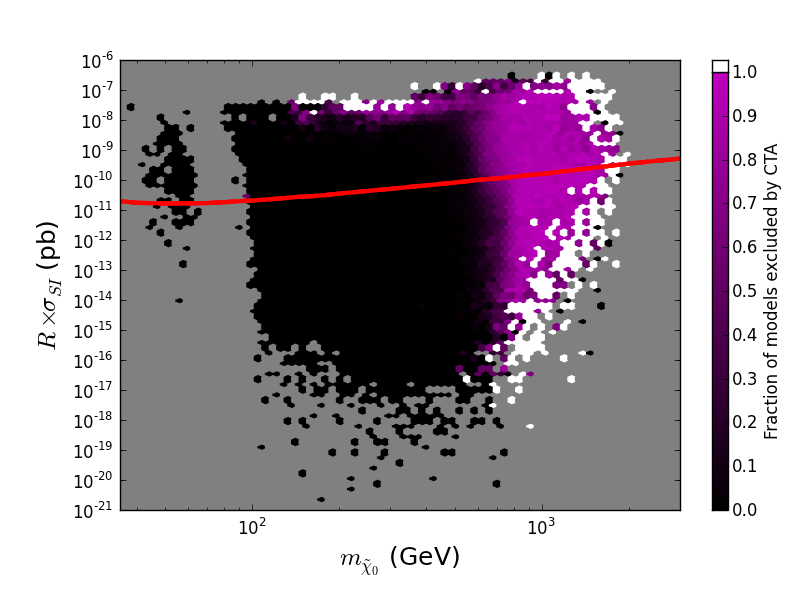}
\hspace{-0.50cm}
\includegraphics[width=3.5in]{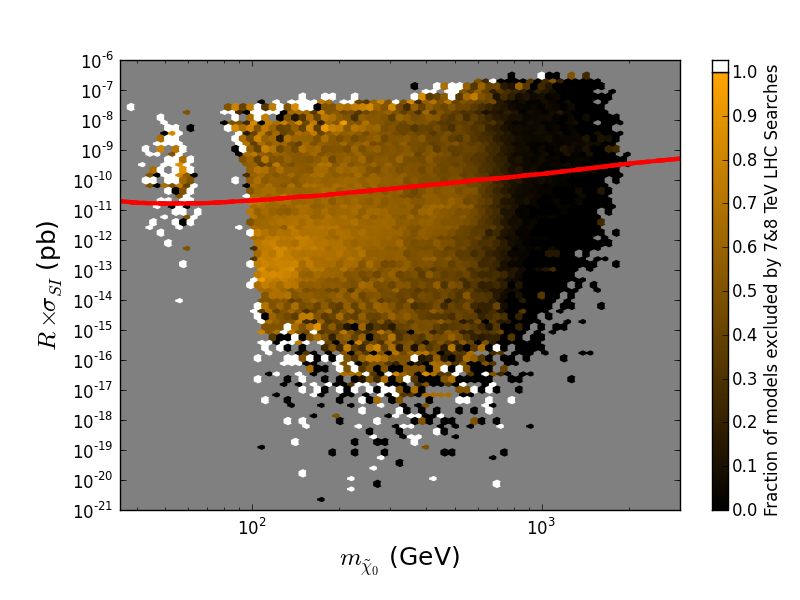}}
\vspace*{0.50cm}
\centerline{\includegraphics[width=3.5in]{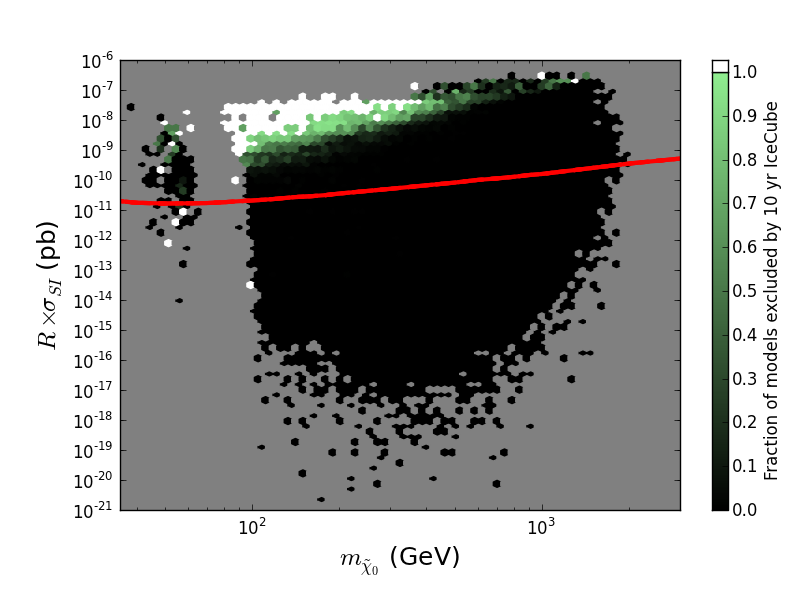}
\hspace{-0.50cm}
\includegraphics[width=3.5in]{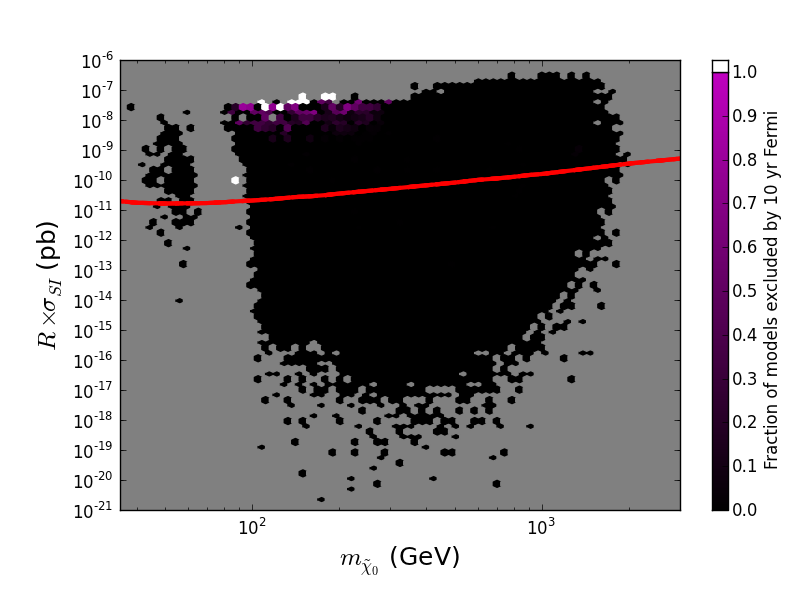}}
\vspace*{-0.10cm}
\caption{Comparisons of the various search capabilities projected into the LSP mass-scaled SI cross section plane, showing the fraction of 
models in each bin which are excluded by CTA (top left), the LHC (top right), IceCube (bottom left) and \Fermi ~(bottom right). The expected 
SI cross-section upper limit from LZ with 1000 days of data is also shown in each case.}
\label{figyy1}
\end{figure}

Fig.~\ref{figyy1} compares the various search capabilities in the LSP mass-scaled SI cross section plane, to which the SI DD searches are 
most directly applicable. Here we see the regions in this plane to which individual searches are most sensitive. In particular, we show the 
fractions of models excluded by CTA (top left), the 7 and 8 TeV LHC searches (top right), IceCube (bottom left) and \Fermi ~(bottom right) 
projected into this plane. In each case, we also show the expected limit from SI LZ, which applies directly to this plane. We see that both 
IceCube and \Fermi ~probe models with low LSP masses and large SI cross sections, where the LSP tends to be a bino-Higgsino admixture; 
this region is also accessible to the DD experiments such as LZ. On the other hand, CTA has access to the heavy LSP region, where there are 
a large fraction of relatively pure winos and Higgsinos, while the LHC coverage is mostly concentrated (for now) on the relatively low 
mass LSP region.

\begin{figure}[htbp]
\centerline{\includegraphics[width=3.5in]{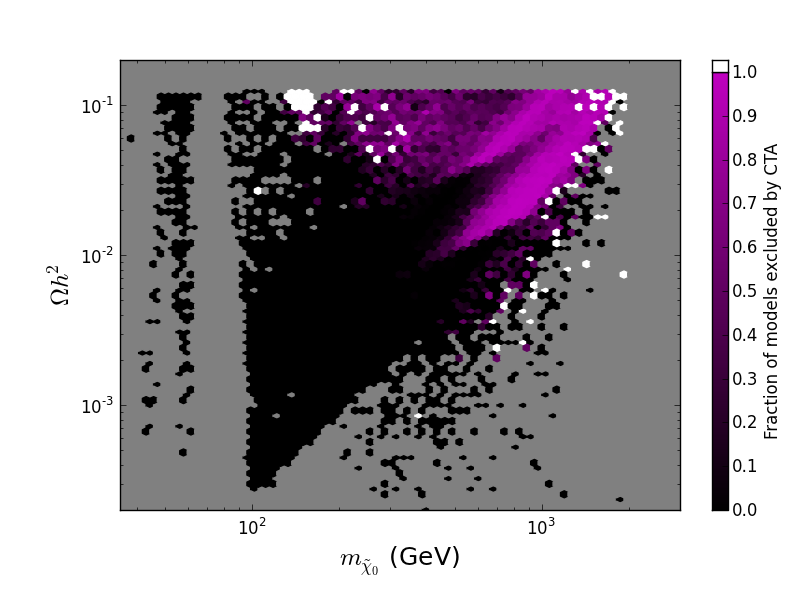}
\hspace{-0.50cm}
\includegraphics[width=3.5in]{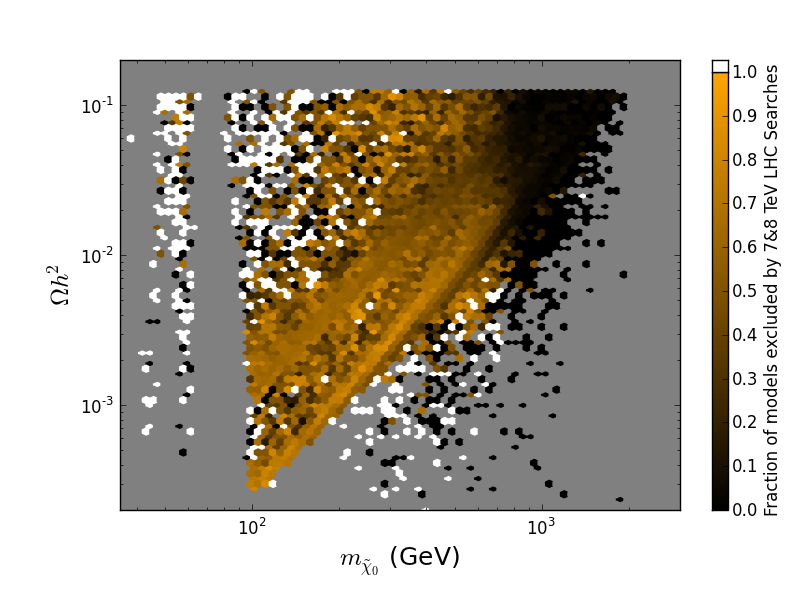}}
\vspace*{0.20cm}
\centerline{\includegraphics[width=3.5in]{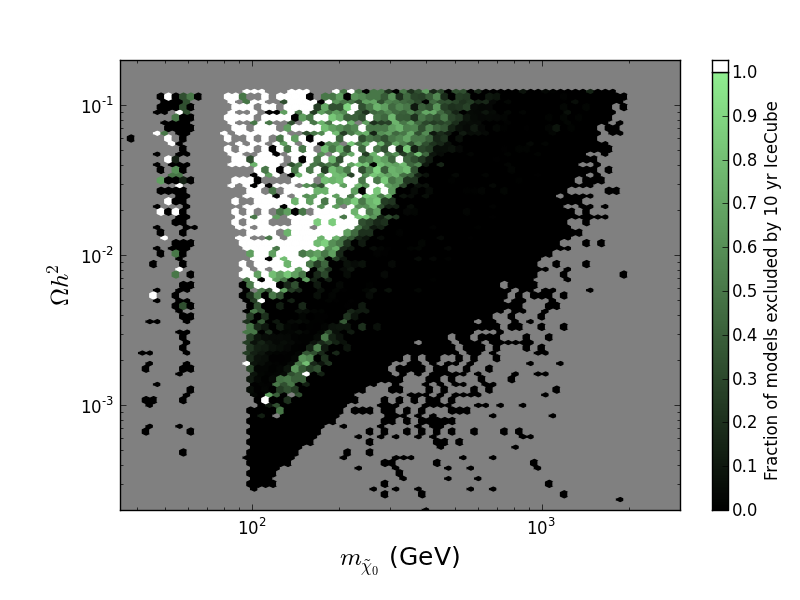}
\hspace{-0.50cm}
\includegraphics[width=3.5in]{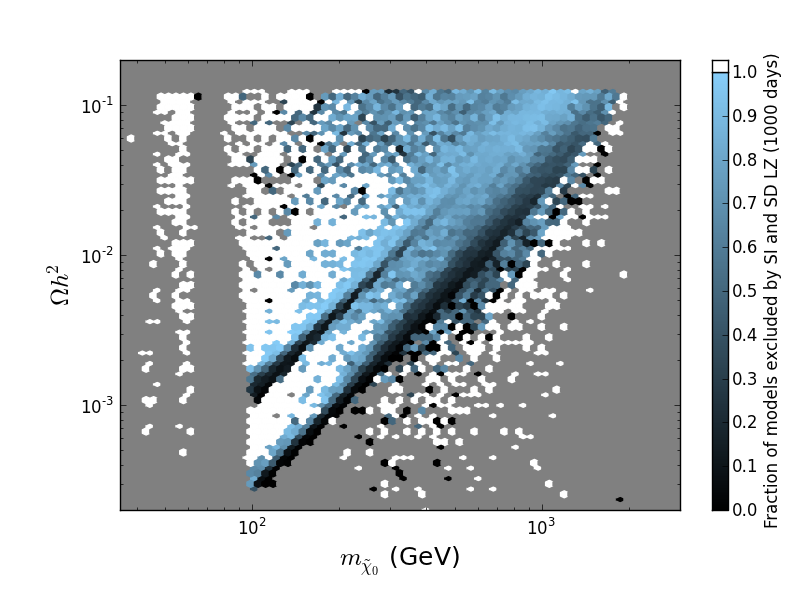}}
\vspace*{0.20cm}
\centerline{\includegraphics[width=3.5in]{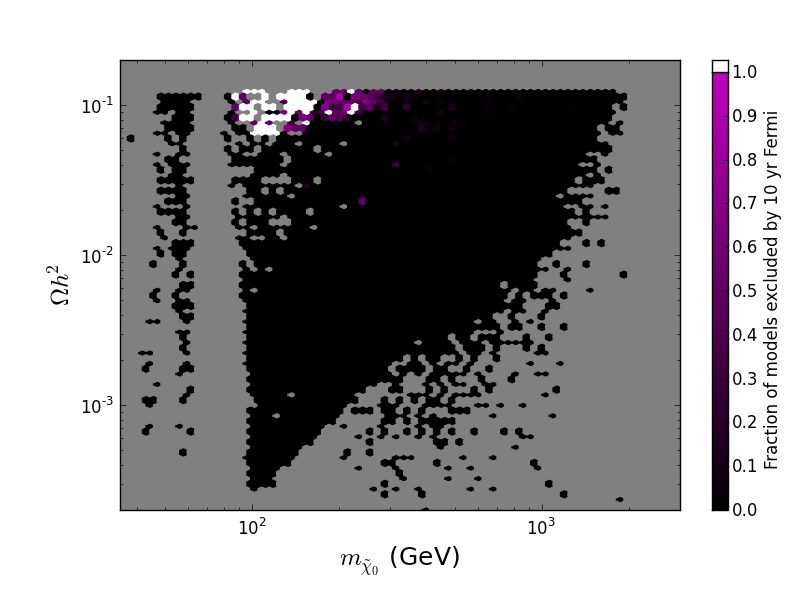}}
\vspace*{-0.10cm}
\caption{Comparisons of the various search capabilities in the LSP mass-relic density plane, showing the fraction of models excluded by 
CTA (top left), the LHC (top right), IceCube (middle left), LZ (middle right) and \Fermi ~(bottom).}
\label{figyy3}
\end{figure}

We now can project these same searches into the thermal relic density-LSP mass plane which provides a different aspect to the coverage of 
the model space by the various searches. In Fig.~\ref{figyy3} we display the fractions of models excluded by CTA, the 7 and 8 TeV LHC searches, 
IceCube, LZ and \Fermi~ in this plane. We again see that the regions excluded by the various experiments overlap significantly; for example, 
many different experiments will be sensitive to the ``well-tempered neutralino'' scenario{\footnote {In the case of an actual DM {\it discovery}, 
the existence of a substantial overlap between the regions of experimental coverage within this parameter space will be very helpful when trying 
to determine the specific nature of the LSP.}}. On the other hand, there are important cases where experiments complement each other to exclude 
a much larger fraction of models than could be seen by any one experiment. One example of this is the sensitivity of CTA to high-mass LSPs 
which are difficult to observe at the LHC; another is the sensitivity of the LHC to relatively light winos, which have a very low relic density 
and are generally missed by both the DD and ID DM searches. We also note several other features of these plots: First, CTA, as expected, does 
an excellent job at excluding most of the models with LSP masses above $\sim 250$ GeV that saturate the relic density. However, for larger masses 
the CTA coverage extends down to relic densities as much as a factor of $\sim 10$ or more below the WMAP/Planck value. \Fermi~ is seen to cover 
only the low LSP mass region with relic densities not far from the thermal value, while IceCube can go to much lower relic densities provided 
the LSP mass is below $\sim 500$ GeV or so. LZ has sensitivity throughout this plane but does best for LSP masses below $\sim 300$ GeV, even 
for models with very low relic densities. Of course, even for LSP masses up to 1-2 TeV, the LZ sensitivity remains reasonably good. As noted 
already, the LHC is presently seen to be effective mainly at lower LSP masses below $\sim 500-600$ GeV. The LHC coverage is relatively uniform 
with respect to the relic density, but of course the fraction of models excluded is very high in the case of very light LSPs.  Extending the 
LHC energy to 14 TeV will substantially improve its ability to find heavy LSPs, as we will see below.

\begin{figure}[htbp]
\centerline{\includegraphics[width=7.0in]{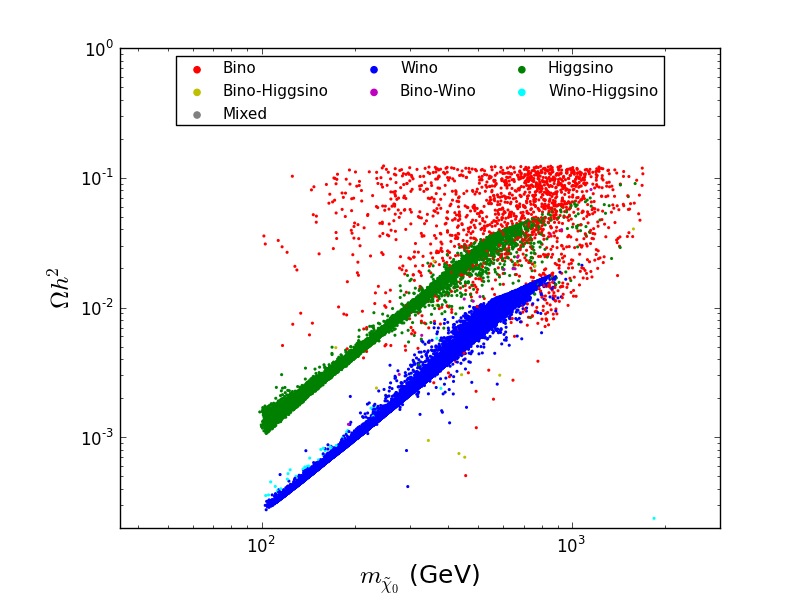}}
\vspace*{-0.10cm}
\caption{Thermal relic density as a function of the LSP mass for all pMSSM models, surviving after all searches, color-coded by the 
electroweak properties of the LSP. Compare with Fig.~\ref{fig00}. }
\label{figzz}
\end{figure}

Fig.~\ref{figzz} shows the impact of combining all of the different searches (in the case of null search results) in this same $\Omega h^2$-LSP 
mass plane, which should be compared with the same plot for the original model set, shown above in Fig.~\ref{fig00}. Here we see that ($i$) the 
models that were in the light $h$ and $Z$-funnel regions have completely evaporated through a combination of the SI and SD LZ analyses, 
($ii$) the well-tempered neutralinos are seen to be completely gone, mostly due to LZ and IceCube, with additional help from \Fermi; ($iii$) 
the possibility of almost pure Higgsino or wino LSPs even approximately saturating the relic density has vanished thanks to CTA and ($iv)$ 
the mixed wino-Higgsino models, due to a combination of measurements, have also completely disappeared. ($v$) The only models remaining 
which {\it do} saturate the WMAP/Planck relic density are those with binos with $A$ resonant or co-annihilations. ($vi$) We find that 
$\sim 75.5\%$ of all the pMSSM models have been excluded by at least one of the searches considered here.

\section{14 TeV LHC Searches and the ILC}

\begin{figure}[htbp]
\centerline{\includegraphics[width=3.5in]{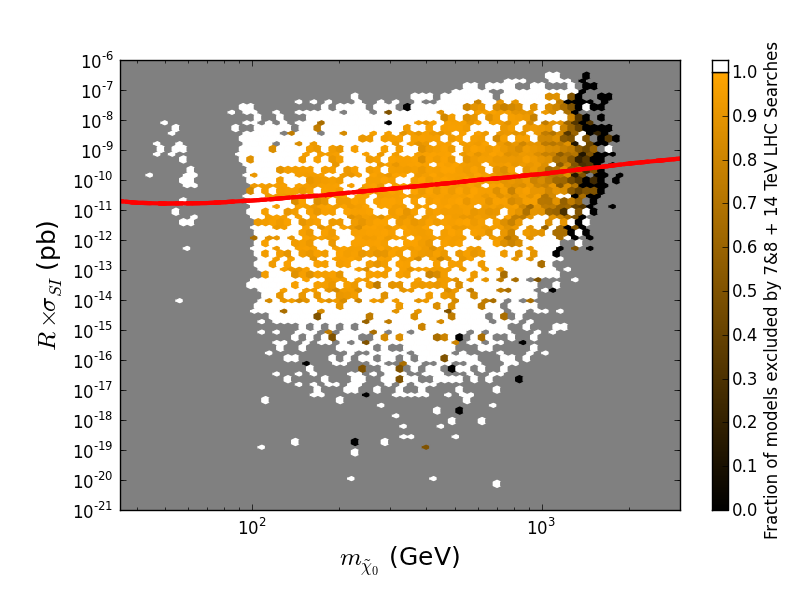}
\hspace{-0.50cm}
\includegraphics[width=3.5in]{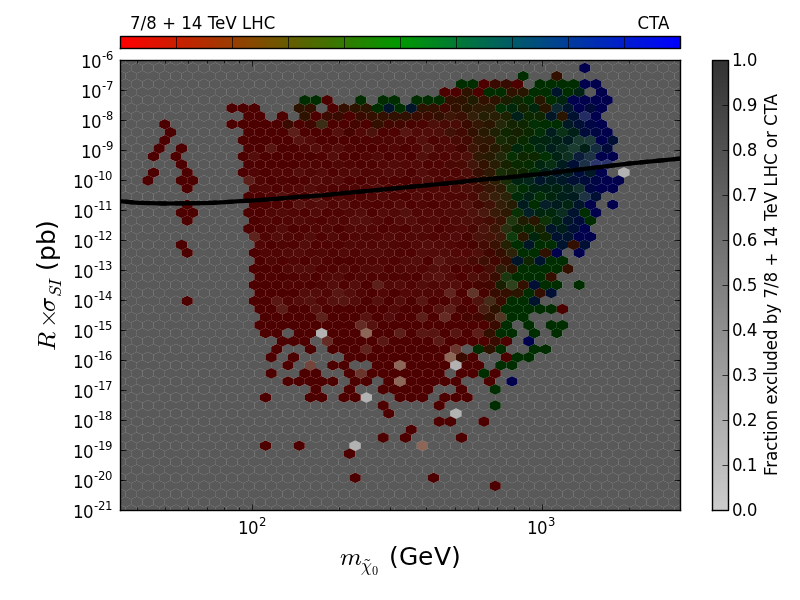}}
\vspace*{-0.10cm}
\caption{The fraction of models which are excluded by the combination of the 14 TeV jets + MET and the 0$\ell$ + 1$\ell$ stop searches 
with 300 fb$^{-1}$, shown in the LSP mass-scaled SI cross section plane (left panel) and a comparison between the fractions of models 
excluded by the LHC and CTA in this plane (right panel). The expected limit on the Xenon SI cross section from LZ is also shown in both cases.}
\label{14TeVComp}
\end{figure}

We consider first the addition of the 14 TeV jets + MET and corresponding 0- and 1-lepton stop searches with 300 fb$^{-1}$ of integrated luminosity 
to the full set of 7+8 TeV searches considered previously (restricting our analysis to the subset of $\sim$ 45k models with $m_h = 126\pm 3$ GeV). 
The left panel of Figure~\ref{14TeVComp} shows the reach of these combined LHC searches in the LSP mass - SI cross-section plane, which is analogous  
to the right panel of Figure~\ref{figyy1}. Note that in both cases the effectiveness of the LHC searches falls off sharply above a particular LSP 
mass, since above this limit the spectrum is generically either too heavy or too compressed to be observed given our scan ranges. The addition of 
the 14 TeV searches effectively doubles this cutoff, from $\sim$ 700 GeV to $\sim$ 1400 GeV, so that most LSPs in our model set can be excluded 
given colored sparticle masses below $\sim$ 2-3 TeV. We also see that the LHC now excludes a very high fraction (but not all) of the models with 
LSPs lighter than the aforementioned cutoff. The large fraction of models which are excluded by the 14 TeV data is unsurprising, since our chosen 
upper limits for the sparticle masses were designed to ensure that most models would be (at least kinematically) accessible at the 14 TeV LHC.

The right panel of Figure~\ref{14TeVComp} compares the reaches of the LHC and CTA in this plane, analogous to the right panel of 
Figure~\ref{figxx}. We see that the LHC and CTA sensitivities now exhibit a sizable region of overlap. However, the blue region on the far 
right edge of the plot shows that CTA will still be sensitive to LSP masses beyond the reach of the 14 TeV LHC. We also note that LSPs heavy 
enough to be seen by CTA are generally too heavy to be seen in direct (\eg, monojet) LHC searches, meaning that the LHC is excluding these models 
by observing other (mostly colored) sparticles. It is therefore important to note that CTA has the potential to exclude winos and Higgsinos 
with nearly thermal relic densities  \textit{regardless} of the characteristics of the rest of the sparticle spectrum. Of course, the improved 
reach of the LHC means that there is also an increasing overlap between the LHC and LZ, however we see that (as before) the LHC searches are 
mostly independent of the SI cross-section (the white areas around the edges of the plot generally result from low statistics in those regions, 
increasing the likelihood that all of the models in a given bin will be excluded). Of course, from a discovery perspective, the increased 
overlap means that there is more potential for a signal to be observed by two, or even all three, experiments, which would greatly aid in 
characterizing the LSP and model properties. Finally, we note that not only has the total fraction of models excluded by the combined experiments 
increased dramatically (from 75.5\% to 98.0\%) as a result of the improved LHC reach, but the fraction of models not seen by the LHC 
which are excluded by direct or indirect detection has increased (from 54.8\% to 59.3\%), since more of the undetected models have heavy LSPs 
and are therefore likely to be excluded by CTA.

Assuming the discovery of a light DM candidate at the LHC the ILC will have an important role to play. To fully complete the picture of our 
understanding of neutralino (or mixed) DM we will need to know the electroweak properties of the LSP, something which is expected to be quite 
difficult to do at the LHC in most of the model parameter space even with large integrated luminosities. However, such determinations can be 
quite easily performed in a number of ways with fairly high precision at the ILC assuming that the LSP is kinematically accessible. The most 
obvious techniques for doing this are via the production and decay of the heavier electroweakinos and/or sleptons or by `monophoton plus 
invisible' cross section measurements.

\section{Conclusions}

These results lead to a number of interesting conclusions:

\begin{itemize}

\item{Even if the LSP {\it does not} make up all of the DM, it can still be observed in both direct and indirect detection experiments as 
well as neutrino experiments such as IceCube. Of course, searches at the LHC are not influenced by the LSP relic density.}

\item{The set of models remaining after all the searches are performed, yielding null results,  that saturate the thermal relic density consist 
almost uniquely of those with (co)annihilating bino LSPs.}

\item{SI direct detection via LZ, CTA, and the LHC do most of the heavy lifting in terms of complementary searches covering the pMSSM parameter space.}

\item{Multiple/overlapping searches allow for extensive parameter space coverage which will be of particular importance if a DM signal is observed.}

\item{Most of the experiments are seen to provide complementary probes of the pMSSM parameter space.} 

\item{The strength of the LHC component in these searches increases significantly with the inclusion of the 14 TeV LHC searches. However, these 
searches dependent on the rest of the model spectrum and therefore do not provide complete coverage of any given LSP scenario, in 
contrast to DM searches which rely more directly on the LSP properties.}

\item{The ILC has an important role to play in determining the electroweak properties of the the LSP after discovery to complete the whole DM picture.}

\end{itemize}

In summary, the pMSSM provides an excellent tool for studying complementarity between different approaches to the search for and exploration of dark 
matter. Hopefully, DM will soon be discovered so that we can employ the complementary probes discussed above to ascertain its true nature.

\section*{Acknowledgments}

The author would like to thank his collaborators M.Cahill-Rowley, R. Cotta, A. Drlica-Wagner, S. Funk, J. Hewett, A. Ismail and M. Wood for making 
this analysis possible. This work was supported by the Department of Energy, Contract DE-AC02-76SF00515.

\end{document}